\documentclass{iau}
\usepackage{graphicx}

\title[IAUS291.~~Radio counterparts of gamma-ray pulsars] 
{Radio counterparts of gamma-ray pulsars} 

\author[L. Guillemot et al.]  
{L. Guillemot$^1$, on behalf of the \textit{Fermi} LAT Collaboration, \\
the \textit{Fermi} Pulsar Search Consortium and \\ the \textit{Fermi} Pulsar Timing Consortium}

\affiliation{$^1$Max-Planck-Institut f\"ur Radioastronomie, Auf dem H\"ugel 69, 53121 Bonn, Germany \\ 
email: {\tt guillemo@mpifr-bonn.mpg.de} \\[\affilskip]
}

\pubyear{2012}
\volume{291}  
\jname{\mbox{Neutron Stars and Pulsars: Challenges and Opportunities after 80 years}}
\editors{J. van Leeuwen, ed.} 
\begin{document}

\maketitle

\begin{abstract}

Observations of pulsars with the Large Area Telescope (LAT) on the \textit{Fermi} satellite have revolutionized our view of the gamma-ray pulsar population. For the first time, a large number of young gamma-ray pulsars have been discovered in blind searches of the LAT data. More generally, the LAT has discovered many new gamma-ray sources whose properties suggest that they are powered by unknown pulsars. Radio observations of gamma-ray sources have been key to the success of pulsar studies with the LAT. For example, radio observations of LAT-discovered pulsars provide constraints on the relative beaming fractions, which are crucial for pulsar population studies. Also, radio searches of LAT sources with no known counterparts have been very efficient, with the discovery of over forty millisecond pulsars. I review radio follow-up studies of LAT-discovered pulsars and unidentified sources, and discuss some of the implications of the results.

\end{abstract}


\firstsection 
\section{Introduction}

The Large Area Telescope (LAT), a pair conversion telescope sensitive to gamma-ray photons with energies between 20 MeV and more than 300 GeV, is the primary instrument on the \textit{Fermi} observatory launched in June 2008. The LAT is the most sensitive GeV telescope to have ever flown, it has a large field of view of 2.4 sr, and it operates in a continuous sky survey mode so that it covers the entire sky every two orbits ($\sim$ 3 hours). These characteristics make the LAT an ideal instrument for studying pulsars in the GeV domain, and they have also led this instrument to provide the most complete picture of the gamma-ray sky to date. 

GeV pulsar observations have been one of the main successes of the
LAT. After three years of LAT data taking, the number of pulsars known
to emit in gamma rays has increased from less than ten objects
(e.g. \cite[Thompson 2001]{Thompson2001}) to 117 pulsars detected with high
significance\footnote{A public list of LAT-detected pulsars is
  available at:\\
  {\tt https://confluence.slac.stanford.edu/display/SCIGRPS/Detected+Gamma-Ray+Pulsars} \\ The
  Second \textit{Fermi} Large Area Telescope Catalog of gamma-ray
  pulsars, currently in preparation within the \textit{Fermi}
  collaboration, will list the properties of the 117 pulsars detected
  in the first three years of LAT operation.}. Gamma-ray pulsars are
currently roughly split into three categories of comparable size: one
third are ``normal'' radio pulsars, another third are ``normal''
pulsars discovered through blind searches of the LAT data (see Saz
Parkinson, these proceedings, for a review of blind search techniques
and main results), and a third are radio ``millisecond'' pulsars
(MSPs). Most of these pulsars have been detected in the gamma-ray data
using ephemerides obtained from radio timing observations, conducted
over several months to several years (\cite[Smith et al.\ 2008]{Smith2008}). 

Radio pulsar observations in support of the \textit{Fermi} mission have also been useful for determining which of the LAT-discovered pulsars also emit in radio, which provides important information about the geometry of emission in the pulsar magnetosphere, and also for determining the nature of certain \textit{Fermi} LAT sources with no known associations (so-called ``unassociated sources''). Radio pulsar searches in unassociated LAT sources have indeed led to the discovery of many new pulsars, including a large fraction of radio MSPs in binary orbits. 

In these proceedings I review radio observations of pulsars discovered in blind searches of the LAT data and searches for radio pulsars at the position of LAT unassociated sources, and discuss some implications of the results in terms of radio and gamma-ray beaming of pulsars.

\section{Radio detections of blind search pulsar discoveries}

A total of 36 pulsars have so far been discovered in blind searches of
the LAT data
(\cite[Abdo et al.\ 2009, Saz Parkinson et al.\ 2010, Saz Parkinson
  2011, Pletsch et al.\ 2012a, Pletsch et al.\ 2012b]{FermiBlindSearch,SazParkinson2010,SazParkinson2011,Pletsch2012a,Pletsch2012b}). Following
these discoveries, systematic searches for radio pulsations from these
pulsars have been conducted at several large radio telescopes around
the world (see e.g. \cite[Ray et al.\ 2011]{Ray2011}). Detections or non-detections of radio emission from gamma-ray pulsars provide information on the emission geometry in their magnetosphere, and also useful constraints on the radio and gamma-ray beaming fractions, which are critical for population studies. The searches for radio emission resulted in the detection of four of the LAT-discovered pulsars (\cite[Camilo et al.\ 2009, Abdo et al.\ 2010, Pletsch
et al.\ 2012a]{Camilo2009,FermiJ1907,Pletsch2012a}). A tentative detection of 10 aligned sub-pulses from a fifth pulsar at 34.5 MHz was reported by \cite{Maan2012}, but additional searches are needed due to the low statistical significance of the result. Stringent upper limits on the radio fluxes of the other pulsars were also placed. 

\begin{figure}[htpb]
\begin{center}
\includegraphics[width=3.1in]{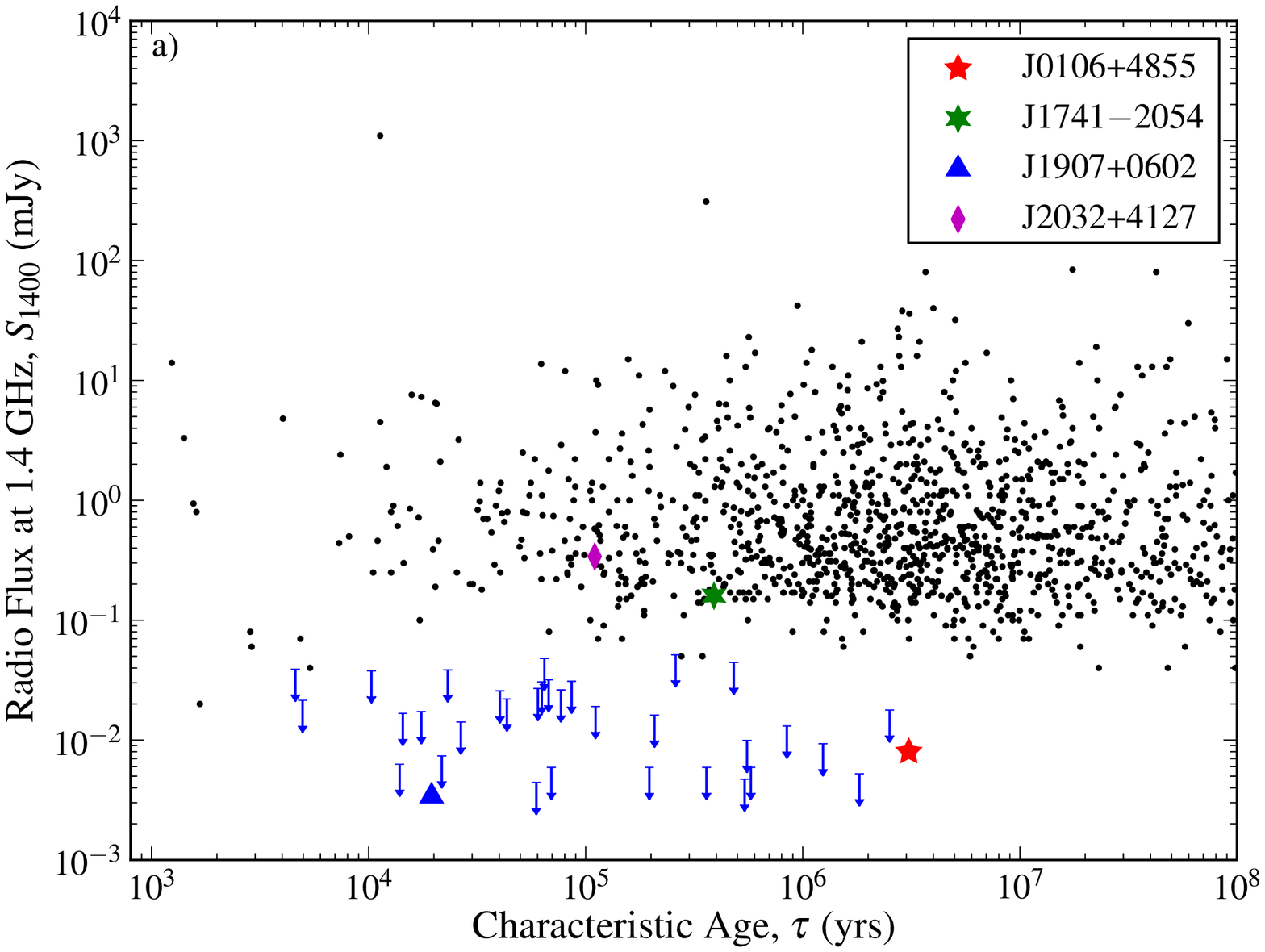} 
\includegraphics[width=3.1in]{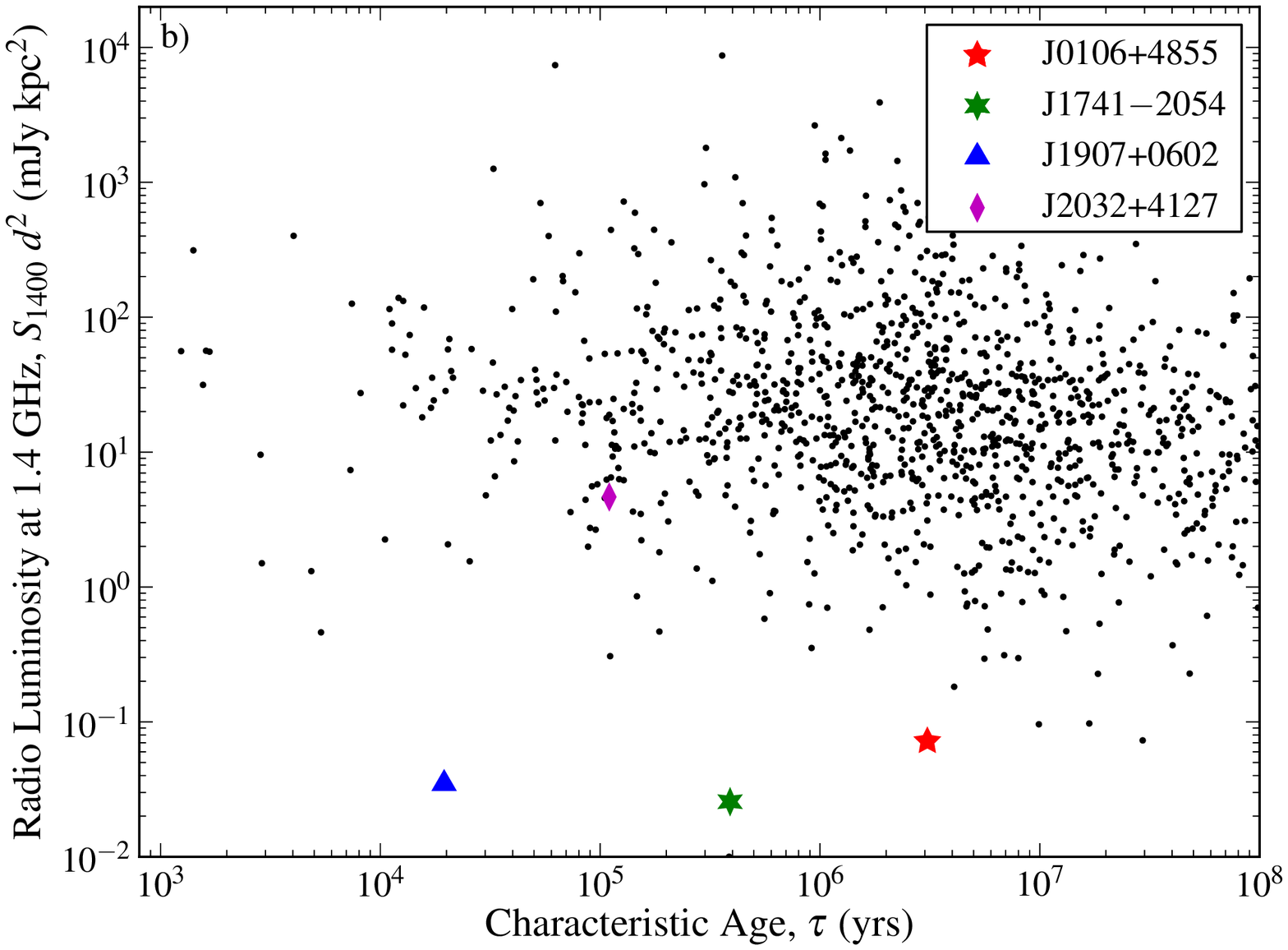} 
\caption{\textit{Top:} radio flux density at 1.4 GHz, $S_{1400}$, as a function of the characteristic age, $\tau = P / (2 \dot P)$ for the normal population of pulsars. Black dots represent radio pulsars from the ATNF catalog (\cite[Manchester et al.\ 2005]{ATNF}), and colored symbols represent pulsars discovered in blind searches of \textit{Fermi} LAT data, later detected in the radio domain. Upper limits on the radio flux density of other LAT blind searches pulsars are shown as blue arrows. \textit{Bottom:} radio pseudo-luminosity, $S_{1400} d^2$, as a function of $\tau$, for the same pulsar sample.} 
\label{radio}
\end{center}
\end{figure}

The fluxes of the radio-detected LAT-discovered pulsars scaled to 1400 MHz are displayed in Figure~\ref{radio}a. Two of them, PSRs J1741$-$2054 and J2032+4127, have fluxes that are relatively common among other normal radio pulsars. On the other hand, PSRs J0106+4855 and J1907+0602 have extremely low values of the radio flux, similar to those of the faintest previously-known pulsars. The upper limits on the radio fluxes of the other new gamma-ray pulsars, also shown in Figure~\ref{radio}a, indicate that their radio fluxes are at most in the same order as those of PSRs J0106+4855 and J1907+0602. It is clear that \textit{Fermi} made the discovery of the 36 blind search pulsars significantly easier. 

For pulsars detected in radio, one can determine the dispersion measure (DM), corresponding to the quantity of free electrons along the line-of-sight, which in turn can be used to determine distances to the pulsars, via a model for the distribution of free electrons in the Galaxy (e.g. NE2001; see \cite[Cordes \& Lazio 2002]{NE2001}). In turn, these distance estimates can be used to derive radio or gamma-ray pseudo-luminosities for these pulsars, which is impossible in the absence of any distance information. Figure~\ref{radio}b shows the pseudo-luminosities of the LAT-discovered pulsars detected in radio. As can be seen from this plot, three of the four radio-detected pulsars have extremely low radio pseudo-luminosities compared to the rest of the normal pulsar population. Understanding the causes of non-radio-detections of the majority of the LAT-discovered pulsars as well as understanding the very low luminosities of the radio-detected ones will be of paramount importance for pulsar population studies.

\section{Radio MSP discoveries in \textit{Fermi} sources}

In addition to searches for radio emission from LAT-discovered
pulsars, radio telescopes have also conducted deep searches for
pulsations from \textit{Fermi} LAT sources with no known
associations. These unassociated sources are numerous: they represent
$\sim$ 30\% of the 1451 sources listed in the \textit{Fermi} LAT
Second Source Catalog (2FGL; \cite[Nolan et al.\ 2012]{Fermi2FGL}),
and among this source category, a large fraction have gamma-ray
emission properties reminiscent of those of known pulsars, i.e., low
gamma-ray flux variabilities and significant curvature in their
emission spectra (cf. Figure~17 of \cite[Nolan et al.\ 2012]{Fermi2FGL}), in contrast with e.g. blazars and other active galactic nuclei (see e.g. \cite{Lee2012} or \cite{FermiClassification} for examples of \textit{Fermi} LAT source classification studies). Radio pulsar searches of \textit{Fermi} LAT sources therefore offered the possibility to reveal the nature of some of the LAT unassociated sources with pulsar properties. 

Searches for pulsars in unidentified high-energy sources had already
been conducted prior to the \textit{Fermi} mission: for instance,
pulsars have been searched in EGRET sources, with limited success as
searches were complicated by the large number of radio pointings
required to cover a typical gamma-ray source (see for instance
\cite[Champion et al.\ 2005]{Champion2005} or \cite[Crawford et al.\ 2006]{Crawford2006}). In contrast, \emph{Fermi} LAT sources are typically localized to within 10$'$ due to the improved angular resolution (\cite[Nolan et al.\ 2012]{Fermi2FGL}). Large radio telescopes having beam widths comparable to this localization accuracy, they can cover \emph{Fermi} LAT sources in only one or a few pointings, making these searches easier and more efficient. 

\begin{figure}[tpb]
\begin{center}
\includegraphics[width=2.1in,angle=90]{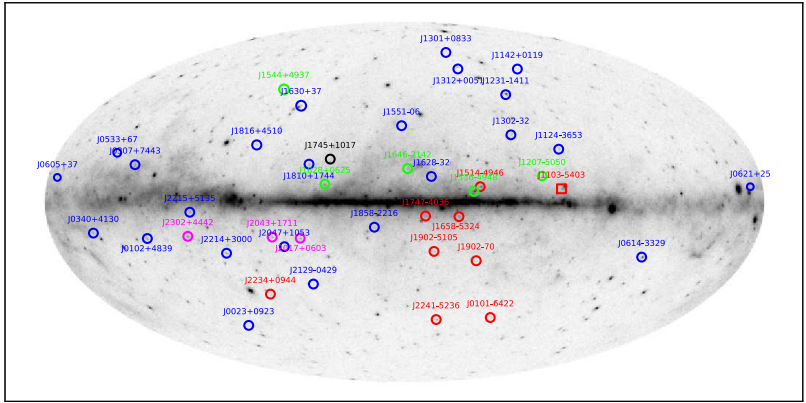} 
\caption{Map of the gamma-ray sky as seen with the \textit{Fermi} LAT
  in Galactic coordinates, with the locations of the 43 radio MSPs
  discovered in searches for pulsars in \textit{Fermi} unassociated
  sources. MSPs discovered at the GBT are shown in blue, while pulsars
  found with the GMRT are in green, Effelsberg in black, Parkes in
  red, and Nan\c cay in magenta. PSR J1103$-$5403 is represented with
  a different symbol as it has been shown not to be associated with
  the LAT source in which it was found (\cite[Keith et al.\ 2011]{Keith2011}).} 
\label{pscmsps}
\end{center}
\end{figure}

Radio pulsar searches in LAT unassociated sources undertaken so far
have led to the discovery of 43 new radio MSPs, and an additional four
normal pulsars (e.g. \cite{Ransom2011}, \cite{Keith2011},
\cite{Guillemot2012}; a full listing of pulsar discoveries in
\textit{Fermi} sources to date with complete references is available
in \cite[Ray et al.\ 2012]{Ray2012}). The locations of MSPs found in these surveys are displayed in Figure~\ref{pscmsps}. It is apparent that the new MSPs are widely distributed in Galactic latitudes, which have been only modestly surveyed in past radio pulsar searches with sensitivity to this type of pulsars. These new MSPs are observed to have somewhat different properties than the previously known MSP population in the Galactic disk: for instance, at least 10 ``black widow'' and four ``redback'' systems have been discovered in \textit{Fermi} LAT sources, while only three black widows and one redback were known in the Galactic disk prior to \textit{Fermi}'s launch (see \cite{Roberts2011} and references therein). Another difference is the period distribution of the new MSPs, observed to be concentrated toward smaller values than those of previously-known Galactic disk MSPs (see Figure 4 of \cite[Ray et al.\ 2012]{Ray2012}). It is thus clear that pulsar searches in \textit{Fermi} LAT sources are biased in a different way than traditional radio surveys, and thus help to get a more complete picture of the population of MSPs in the Galaxy. 

The new MSPs have been timed at radio wavelengths following their discovery, and as they became available the timing parameters obtained from these measurements have been used to phase-fold the LAT data. Because the MSPs were found within the error boxes of LAT sources, these searches unsurprisingly resulted in the detection of gamma-ray pulses in most cases, one counter-example being PSR~J1103$-$5403, discovered coincidentally in a source likely powered by an active galactic nucleus (\cite[Keith et al.\ 2011]{Keith2011}). 

Unlike several of the LAT-discovered pulsars detected in radio, these new MSPs are not particularly faint and would have eventually been found by current or future radio pulsar surveys. However, the \textit{Fermi} LAT accelerated their discovery by guiding radio telescopes to the pulsar-like sources in which they were found, which enabled quicker inclusions of these MSPs in pulsar timing arrays aiming at detecting low frequency gravitational waves (a review of pulsar timing array projects can be found in Hobbs, these proceedings). Furthermore, unassociated sources are often searched for pulsars multiple times, as binary eclipses or bad scintillation states can prevent detection on particular epochs. Such multiple re-observations are generally impossible in standard surveys, which are time-limited. This factor also likely contributed to the many MSP discoveries in \textit{Fermi} LAT unassociated sources.

Contrasting with the many radio MSP discoveries, only four normal
radio pulsars have been discovered within the error boxes of
\emph{Fermi} LAT sources, all of them at low Galactic latitudes. One
of them, PSR~J2030+3641, has been detected as a pulsed gamma-ray
emitter since its discovery (\cite[Camilo et al.\ 2012]{Camilo2012}). As is discussed in \cite{Camilo2012}, this is likely due to the fact that past surveys of the Galactic plane have found most of the detectable normal pulsars. Based on the small number of radio pulsar detections in low Galactic latitude gamma-ray sources, and the fact that most of the LAT-discovered pulsars are undetected in radio, they also proposed that the many unassociated gamma-ray sources with pulsar properties remaining near the plane are indeed pulsars, with little or no radio emission emitted toward the Earth. Continued blind searches are therefore important for revealing the nature of these low Galactic latitude gamma-ray sources.

\section{Radio and gamma-ray beaming}

A large fraction of normal gamma-ray pulsars are radio-quiet. Are
there radio-quiet gamma-ray MSPs? Blind searches for MSPs in the LAT
data are complicated by the fact that most known MSPs are in binary
systems ($\sim$ 80\% of them), unlike normal pulsars, which are almost
all isolated. In the absence of any constraints on the orbital
parameters, blind gamma-ray searches for binary MSPs are intractable
with the current computational resources, as the number of parameters
to be tested is too large. It is nevertheless possible to search for
isolated MSPs in the LAT data, with the same techniques as the ones
used for searching for normal isolated pulsars. Until now, such
searches have failed to find any new isolated gamma-ray MSPs
(e.g. \cite[Pletsch et al.\ 2012a]{Pletsch2012a}). 

The fact that no MSP has so far been found in blind searches of the
LAT data, and that all known gamma-ray MSPs are therefore radio-loud,
suggests that most gamma-ray MSPs must be detectable in
radio. Statistics of the current population of gamma-ray sources
corroborate this observation: selecting 249 bright sources from the
2FGL catalog, \cite{Romani2012} noted that 12 sources are radio-loud
MSPs, and that only six sources remain unassociated, thanks to all
source identification efforts, including blind pulsar searches in
radio and in gamma rays. One can draw the following conclusion: even
if all of the six unassociated sources are radio-quiet MSPs, then the
majority of gamma-ray MSPs in this sample are radio-loud. Among these
six unassociated sources, the high Galactic latitude objects
2FGL~J1311.7$-$3429 (\cite[Romani 2012, Kataoka et al.\ 2012]{Romani2012,Kataoka2012}) and
J2339.6$-$0532 (\cite[Romani 2011]{Romani2011}, \cite[Kong et al.\ 2012]{Kong2012}) have clear pulsar-like gamma-ray properties and could therefore host radio-quiet MSPs. Multi-wavelengths studies of these objects have led to the discovery of optical and X-ray counterparts, exhibiting strong modulations of their emission with periods of $\sim$1.56-hr and $\sim$4.63-hr respectively. These systems are likely to be black widow-type MSPs. The constraints on the orbital parameters as well as the precise positions of these objects obtained from these measurements could make blind gamma-ray pulsation searches feasible for these objects (blind gamma-ray search efforts of such systems are presented in Belfiore, these proceedings).

Based on the same sample of 249 bright 2FGL sources, \cite{Romani2012} showed that most normal gamma-ray pulsars selected are radio-quiet. Unlike MSPs, normal pulsars thus seem to have smaller radio beams than gamma-ray beams. Nevertheless, as was shown by \cite{Ravi2010}, the most energetic objects among normal gamma-ray pulsars are all radio-loud, while the fraction of radio-loud pulsars rapidly decreases for less energetic objects. This suggests an interesting correlation between the size of the radio beam and the pulsar properties. Such constraints on the respective sizes of radio and gamma-ray beams for the different pulsar families provide key input for pulsar population studies, underlining the importance of joint radio and gamma-ray observations of pulsars.

\section*{Acknowledgements}

The $Fermi$ LAT Collaboration acknowledges support from a number of agencies and institutes for both development and the operation of the LAT as well as scientific data analysis. These include NASA and DOE in the United States, CEA/Irfu and IN2P3/CNRS in France, ASI and INFN in Italy, MEXT, KEK, and JAXA in Japan, and the K.~A.~Wallenberg Foundation, the Swedish Research Council and the National Space Board in Sweden. Additional support from INAF in Italy and CNES in France for science analysis during the operations phase is also gratefully acknowledged.


\begin{thebibliography}{}

\bibitem[Abdo et al. (2009)]{FermiBlindSearch}{Abdo, A.~A., Ackermann, M., Ajello, M., et al.} 2009,  \textit{Science}, 325,  840 

\bibitem[Abdo et al. (2010)]{FermiJ1907}{Abdo, A.~A., Ackermann, M., Ajello, M., et al.} 2010,  \textit{ApJ}, 711,  64 

\bibitem[Ackermann et al. (2012)]{FermiClassification}{Ackermann, M., Ajello, M., Allafort, A., et al.} 2012,  \textit{ApJ}, 753,  83 

\bibitem[Camilo et al. (2009)]{Camilo2009}{Camilo, F., Ray, P.~S., Ransom, S.~M., et al.} 2009,  \textit{ApJ}, 705,  1 

\bibitem[Camilo et al. (2012)]{Camilo2012}{Camilo, F., Kerr, M., Ray, P.~S., et al.} 2012,  \textit{ApJ}, 746,  39 

\bibitem[Champion et al. (2005)]{Champion2005}{Champion, D.~ÊJ., McLaughlin, M.~A., \& Lorimer, D.~R.} 2005, \textit{MNRAS}, 364, 1011

\bibitem[Cordes \& Lazio (2002)]{NE2001}{Cordes, J.~M., \& Lazio, T.~J.~W.} 2002,  arXiv:0207156 

\bibitem[Crawford et al. (2006)]{Crawford2006}{Crawford, F., Roberts, M.~S.~E., Hessels, J.~W.~T., et al.} 2006,  \textit{ApJ}, 652,  1499

\bibitem[Guillemot et al. (2012)]{Guillemot2012}{Guillemot, L., Freire, P.~C.~C., Cognard, I., et al.} 2012,  \textit{MNRAS}, 422,  1294 

\bibitem[Kataoka et al. (2012)]{Kataoka2012}{Kataoka, J., Yatsu, Y., Kawai, N., et al.} 2012, \textit{ApJ}, 757, 176

\bibitem[Keith et al. (2011)]{Keith2011}{Keith, M.~J., Johnston, S., Ray, P.~S., et al.} 2011,  \textit{MNRAS}, 414,  1292 

\bibitem[Kong et al. (2012)]{Kong2012}{Kong, A.~K.~H., Huang, R.~H.~H., Cheng, K.~S., et al.} 2012, \textit{ApJL}, 747, 3

\bibitem[Lee et al. (2012)]{Lee2012}{Lee, K.~J., Guillemot, L., Yue, Y.~L., Kramer, M., \& Champion, D.~J.} 2012,  \textit{MNRAS}, 424,  2832 

\bibitem[Maan et al. (2012)]{Maan2012}{Maan, Y., Aswathappa, H.~A., \& Deshpande, A.~A.} 2012,  \textit{MNRAS}, 425,  2 

\bibitem[Manchester et al. (2005)]{ATNF}{Manchester, Hobbs, Teoh, \& Hobbs} 2005,  \textit{AJ}, 129,  1993 

\bibitem[Nolan et al. (2012)]{Fermi2FGL}{Nolan, P.~L., Abdo, A.~A., Ackermann, M., et al.} 2012,  \textit{ApJS}, 199,  31 

\bibitem[Pletsch et al. (2012a)]{Pletsch2012a}{Pletsch, H.~J., Guillemot, L., Allen, B., et al.} 2012,  \textit{ApJ}, 744,  105 

 \bibitem[Pletsch et al. (2012b)]{Pletsch2012b}{Pletsch, H.~J., Guillemot, L., Allen, B., et al.} 2012,  \textit{ApJL}, 755,  20 

\bibitem[Ransom et al. (2011)]{Ransom2011}{Ransom, S.~M., Ray, P.~S., Camilo, F., et al.} 2011,  \textit{ApJL}, 727,  16 

\bibitem[Ravi et al. (2010)]{Ravi2010}{Ravi, V., Manchester, R.~N., \& Hobbs, G.} 2010, \textit{ApJL}, 716, 85

\bibitem[Ray et al. (2011)]{Ray2011}{Ray, P.~S., Kerr, M., Parent, D., et al.} 2011,  \textit{ApJS}, 194,  17 

\bibitem[Ray et al. (2012)]{Ray2012}{Ray, P.~S., Abdo, A.~A., Parent, D., et al.} 2012, Proceedings of the 2011 Fermi Symposium, arXiv:1205.3089 

\bibitem[Roberts (2011)]{Roberts2011}{Roberts, M.~S.~E.} 2011, Proceedings of Pulsar Conference 2010, 1357,  127 

\bibitem[Romani (2011)]{Romani2011}{Romani, R.~W.} 2011, \textit{ApJL}, 743, 26

\bibitem[Romani (2012)]{Romani2012}{Romani, R.~W.} 2012,  \textit{ApJL}, 754,  25 

\bibitem[Saz Parkinson et al. (2010)]{SazParkinson2010}{Saz Parkinson, P.~M., Dormody, M., Ziegler, M., et al.} 2010,  \textit{ApJ}, 725,  571 

\bibitem[Saz Parkinson (2011)]{SazParkinson2011}{Saz Parkinson, P.~M.} 2011, Proceedings of Pulsar Conference 2010, arXiv:1101.3096 

\bibitem[Smith et al. (2008)]{Smith2008}{Smith, D.~A., Guillemot, L., Camilo, F., et al.} 2008,  \textit{A\&A}, 492,  923 

\bibitem[Thompson (2001)]{Thompson2001}{Thompson, D.~J.} 2001, \textit{AIPC}, 558,  103 

\end{thebibliography}
\end{document}